\documentclass[9pt,twocolumn,twoside,nolineno]{pnas-new}

\templatetype{pnasresearcharticle} 
\usepackage[separate-uncertainty=true, parse-numbers=false]{siunitx}
\usepackage{comment}
\graphicspath{{Figures/}}

\usepackage{lineno}
\begin{document}

\title{Contactless micro-elastography of single cells using oscillating  microbubbles as shear wave sources}

\author[a,b]{Gabrielle Laloy-Borgna}
\author[a,c]{Maxime Fauconnier}
\author[a]{Sibylle Grégoire}
\author[a,1,2]{Stefan Catheline}
\author[a,1]{Claude Inserra}

\affil[a]{LabTAU, INSERM, Centre Léon Bérard, Université Lyon 1, F-69003, LYON, France}
\affil[b]{Department of Imaging Physics, Delft University of Technology, the Netherlands}
\affil[c]{Medical Ultrasonics Laboratory (MEDUSA), Department of Neuroscience and Biomedical Engineering, Aalto University, Espoo, Finland.}

\leadauthor{Laloy-Borgna}

\significancestatement{Cell mechanics are central to processes such as differentiation, migration, and disease progression, but current imaging tools are often slow or limited to unusually large cells. By harnessing acoustically oscillating microbubbles as localized wave sources, we achieve contactless micro-elastography of cells five times smaller than previously possible, with acquisition times under one millisecond. This advance enables real-time, label-free monitoring of fast biological processes in diverse cell types, offering powerful opportunities for cell biology, mechanopathology, and drug discovery.}

\authorcontributions{GLB, MF, SC and CI designed the experiments. GLB and CI performed the experiments. GLB and MF processed the data. GLB, SC and CI wrote the paper.}
\authordeclaration{The authors do not have any competing interest to declare.}
\correspondingauthor{\textsuperscript{2}To whom correspondence should be addressed. E-mail: stefan.catheline@inserm.fr}
\equalauthors{\textsuperscript{1}C.I.(Claude Inserra) contributed equally to this work with S.C. (Stefan Catheline).}


\begin{abstract}
The mechanical properties of cells play key roles in their physiology, function, physiological and pathological transformations. Micro-elastography has recently emerged as a promising tool to estimate cellular viscoelastic properties within a millisecond, without the need for mechanical modeling. Here, we report a fully contactless approach to single-cell micro-elastography, using acoustically oscillating gas microbubbles positioned near individual cells (20~\textmu m diameter megakaryocytes) as localized shear wave sources. Using this approach, we successfully performed micro-elastography on cells up to five times smaller than those studied in previous works, establishing the smallest single-cell elastography measurements to date. Spherical or non-spherical bubble oscillations generated 15~kHz elastic waves, which we detected using a high-speed camera coupled to a standard bright-field microscope. Noise correlation elastography enabled the measurement of average and local shear-wave velocities within single cells. Our results demonstrate that this method is robust and reproducible across multiple cells from the same cell line, paving the way for real-time, label-free mechanical monitoring of single cells during fast biological processes.
\end{abstract}

\dates{This manuscript was compiled on \today}
\doi{\url{www.pnas.org/cgi/doi/10.1073/pnas.XXXXXXXXXX}}

\maketitle
\thispagestyle{firststyle}
\ifthenelse{\boolean{shortarticle}}{\ifthenelse{\boolean{singlecolumn}}{\abscontentformatted}{\abscontent}}{}

\firstpage[1]{4}


%

\dropcap{T}he mechanical properties of cells are closely linked to their anatomy, function, and pathological state. Factors such as cytoskeletal organization and intracellular water content influence elasticity \cite{charras_animal_2009}, while changes in membrane stiffness during the cell lifecycle regulate molecular uptake \cite{bleil_structure_1980}. Altered mechanical signatures are hallmarks of disease: tumor cells differ mechanically from healthy ones \cite{cross_nanomechanical_2021}, and chemical \cite{grady_cell_2016} and mechanical \cite{laloy-borgna_magnetic_2024} treatments can both alter their elasticity. Monitoring these variations during tumor invasion may help predict metastatic potential \cite{rother_atomic_2014}. Beyond oncology, mechanical properties are also highly informative in reproductive science. In particular, the viscoelastic properties of oocytes correlate with embryo viability \cite{yanez_human_2016}, offering opportunities to optimize embryo selection in in vitro fertilization. These examples illustrate the broad biological and clinical relevance of accurately probing cellular mechanics.

A wide range of methods has been developed to assess single-cell mechanical properties \cite{wu_comparison_2018}, including micropipette aspiration \cite{mitchison_mechanical_1954}, optical tweezers \cite{robertson-anderson_optical_2018}, and indentation techniques such as atomic force microscopy (AFM) \cite{radmacher_molecules_1992,haase_investigating_2015}. While powerful, most approaches require direct contact with the cell, which can introduce artifacts, and typically provide only indirect estimates of elasticity by fitting deformation to rheological models. Moreover, acquisition times range from seconds to hours, preventing the monitoring of rapid biological processes. Brillouin microscopy provides a fully contactless approach to probe cellular mechanical properties by measuring the frequency shift arising from light interaction with thermally induced acoustic fluctuations \cite{prevedel_brillouin_2019}. While it offers subcellular spatial resolution and depth penetration, the mechanical interpretation remains challenging, as the Brillouin shift is linked to the longitudinal modulus and depends on local optical parameters such as refractive index and mass density. 

Optical micro-elastography provides an appealing alternative by generating high-frequency elastic waves inside a cell and detecting them with a high-frame-rate camera. Wave velocity and attenuation are directly related to viscoelastic properties. The method is fast with acquisition times under 1~ms and offers micrometer-scale spatial resolution. However, it has so far relied on micropipettes to generate waves and has been limited to large cells, such as oocytes, which can reach up to \qty{100}{\micro\metre} in diameter \cite{grasland-mongrain_ultrafast_2018,fle_imaging_2023}. The use of holding and vibrating pipettes can also impose local stresses on the cell, leading to local artifacts.

Bubbles have already been considered as shear wave sources in soft solids. Electrolysis-induced bubbles were shown to generate elastic waves in gels \cite{montalescot_electrolysis-induced_2016}, and more recently a bubble undergoing nonspherical collapse was used to measure shear modulus in tissue phantoms \cite{izak_ghasemian_shear_2023}. However, these approaches were either destructive or not compatible with live-cell imaging.

Here, we introduce a fully contactless optical micro-elastography approach compatible with cells of any size. Our method employs an acoustically oscillating gas microbubble positioned near the cell as a localized shear-wave source. The bubble, excited by an ultrasonic transducer, amplifies local velocity fields and efficiently generates mechanical waves within the cell. This enables, for the first time, ultrafast ($< 1~ms$), artifact-free micro-elastography of small cells down to \qty{10}{\micro\metre} in diameter.

\section{Results}
\begin{figure}[ht]
	\centering
	\input{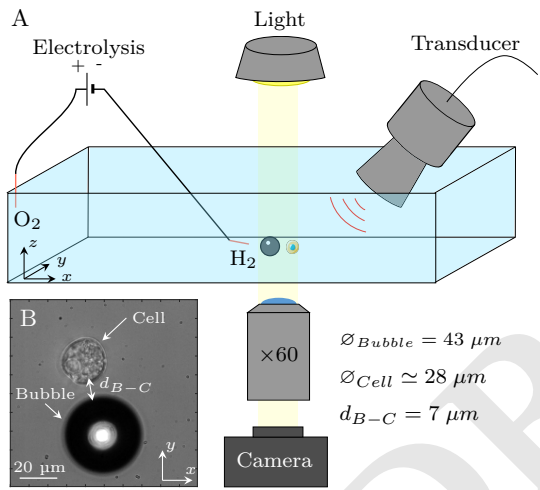}
	\caption{(A) Experimental setup consisting of a water tank filled with cell culture medium placed under an optical microscope with a $\times$60 lens connected to a high-frame-rate camera. A bubble is formed using two electrolysis electrodes, and it is excited using an ultrasonic transducer plunged into the liquid. A cell is placed very close from the bubble. (B) Example of an image recorded by the camera, showing a bubble of \qty{43}{\micro\meter} of diameter and a cell of \qty{28}{\micro\meter} placed \qty{7}{\micro\meter} apart. }
	\label{setup}
\end{figure}

We designed an experiment to observe how a gas microbubble, placed near a single cell, can generate mechanical vibrations inside the cell. The setup is similar to the one previously used for studying the dynamics and microstreaming induced by a substrate-attached microbubble oscillating nonspherically \cite{fauconnier_nonspherical_2020,fauconnier_nonspherical_2022}. It consists in a tank filled with cell culture medium placed under a standard bright-field microscope, as represented in fig.~\ref{setup}A. Bubbles were created through two electrolysis electrodes, then set onto the substrate, before being excited by an ultrasound transducer immersed in the water tank. An exemplary snapshot of the configuration is displayed in fig.~\ref{setup}B, where a bubble and a cell are in the field of view. Cells under interest here were megakaryocytes which are blood cells producing the platelets. A high-speed camera was connected to the microscope and two types of videos were recorded: one with a wide field of view at \qty{6.3}{\kilo fps}, and another one focused on the cell at \qty{120}{\kilo fps}. This allowed to image the bubble-cell ensemble at a low frame rate to analyze the bubble oscillation, but also to obtain a high-frame rate video of the cell only, needed for elastographic analysis. 

\subsection{Bubble oscillations}
The bubble's oscillations were driven by a continuous signal at \qty[parse-numbers=false]{f=30.8}{\kilo\hertz}, corresponding to the resonance frequency of the Langevin transducer. Depending on the applied acoustic pressure, the oscillations could remain purely radial or develop nonspherical components \cite{leighton_acoustic_1994}. Above a certain pressure threshold, parametric instabilities trigger non-spherical deformations of the bubble interface that superimpose on the radial oscillation. The bubble size and driving frequency determine the modal degree n of the nonspherical mode, leading for instance to triangular ($n=3$) or square ($n=4$) contours. The quantification of bubble deformations (fig.~\ref{fig2}A) is based on a Fourier transform of the bubble top-viewed contour \cite{fauconnier_nonspherical_2020}, further discussed in the Materials and Methods section of this document. Repeating the operation for every frame revealed the temporal evolution of the radial and non-spherical components (fig.~\ref{fig2}B). The bubble experiences radial oscillations (red curve) at the driving frequency (\qty{30.8}{\kilo\hertz}) and nonspherical oscillations (green curve) at half the excitation frequency. This subharmonic behavior indicates that the shape oscillations are excited on the first parametric resonance. 

\subsection{Cell vibrations}
\begin{figure*}[!t]
	\centering
	\includegraphics[width=\linewidth]{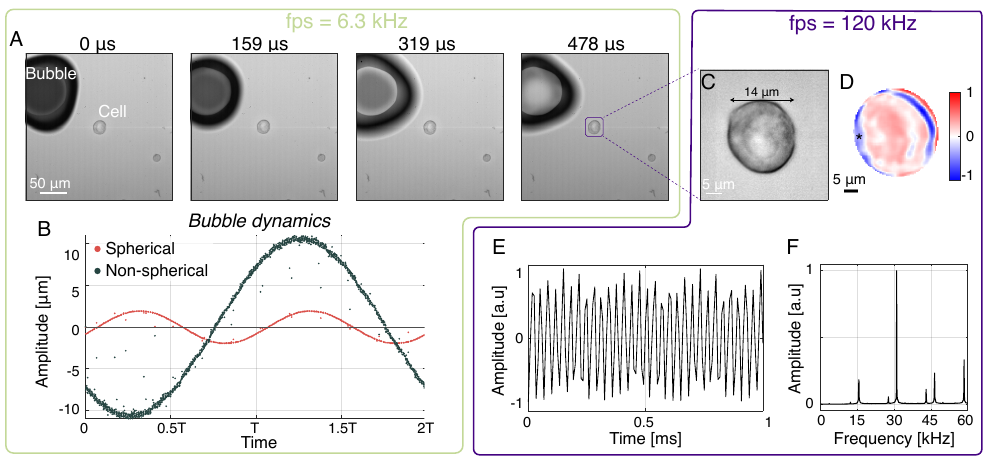}
	\caption{(A) Snapshots of the large field-of-view and low frame rate video showing the bubble and the cell in the same image. (B) Results of the bubble dynamics analysis, showing the temporal dynamics of the spherical (red points) and nonspherical (green points) oscillations. Exemplary snapshot of (C) the narrow field-of-view and higher frame rate video showing only the cell, and (D) of the X-component of the displacement field in the cell. (E) X-component of the displacement field at point * plotted as a function of time and (F) its spectrum.}
	\label{fig2}
\end{figure*}

High-frame-rate recordings (\qty{f_s=120}{\kilo\hertz}) of the cell deformation alone were recorded, and one exemplary snapshot is shown in fig.~\ref{fig2}C. To estimate the local displacements inside the cell, phase-tracking algorithms already described in previous works \cite{laloy-borgna_micro-elastography_2021,laloy-borgna_magnetic_2024,delattre_passive_2021,gregoire_flexural_2024} were applied. The estimated displacement field is shown in fig.~\ref{fig2}D, showing the X-component of the displacement vector at one time point. The signal at point~* (see fig.~\ref{fig2}D) is plotted over time in fig.~\ref{fig2}E, along with its spectrum in fig.~\ref{fig2}F. The peak having the highest amplitude corresponds to the excitation frequency at \SI{30.8}{\kilo\hertz}, while the subharmonic frequency at \SI{15.4}{\kilo\hertz} corresponds to the nonspherical oscillation of the bubble. Both frequency components are retrieved in the cell displacements spectrum. Higher harmonic frequencies are also present (like \SI{45}{\kilo\hertz} and \qty{60}{\kilo\hertz}), together with components higher than \SI{60}{\kilo\hertz}, which appear at 13, 28 and \qty{43}{\kilo\hertz}, due to aliasing (undersampling). At this stage, one first conclusion is that the different frequency components of the bubble dynamics are retrieved in the displacement field measured in the cell.

\subsection{Are the displacements propagating?}
To determine whether the observed displacements correspond to propagating shear waves, we applied temporal band-pass filters centered at 15, 30, and \qty{45}{\kilo\hertz} to the displacement field (Fig.~\ref{fig3}A). $\mathrm{A_{15 kHz}}$ is defined as the 15 kHz peak-amplitude before normalization and is indicated in Fig.~ \ref{fig3}, to be compared with later cases. Filtering around the excitation frequency of \qty{30}{\kilo\hertz} revealed stationary oscillatory patterns without clear propagation, as shown in Movie S2. This is confirmed by the spatio-temporal correlation matrix in Fig.~\ref{fig3}B, computed using noise correlation elastography techniques \cite{laloy-borgna_observation_2023,gregoire_flexural_2024}, which displays stationary, non-propagating features. A similar absence of propagation is observed at \qty{45}{\kilo\hertz} (Movie S3, Fig.~\ref{fig3}C). In contrast, filtering at the subharmonic frequency of \qty{15.4}{\kilo\hertz}—associated with nonspherical bubble oscillations—reveals clear wavefront propagation across the cell. Movie S1 and Fig.~\ref{fig3}E show displacement patterns that persist over multiple frames and shift spatially over time. The corresponding correlation matrix (Fig.~\ref{fig3}D) exhibits oblique streaks spanning the entire cell diameter (\qty{14}{\micro\metre}), indicating shear wave propagation along the X-axis, from right to left, consistent with the visual evidence in Movie S1.

These observations align with the concept of a cutoff frequency, previously introduced in \cite{laloy-borgna_micro-elastography_2021}, which defines the transition above which viscous behavior dominates and shear waves become highly attenuated. In soft solids mimicking biological tissues, this cutoff typically lies below \qty{20}{\kilo\hertz} \cite{laloy-borgna_micro-elastography_2021}. Accordingly, the detection of propagating waves at \qty{15}{\kilo\hertz}—but not at 30 or \qty{45}{\kilo\hertz}—reflects the viscoelastic nature of the cell and confirms the relevance of subharmonic excitation for micro-elastography.

\subsection{Shear wave velocity measurement}
\begin{figure}[!t]
	\centering
	\includegraphics[width=\linewidth]{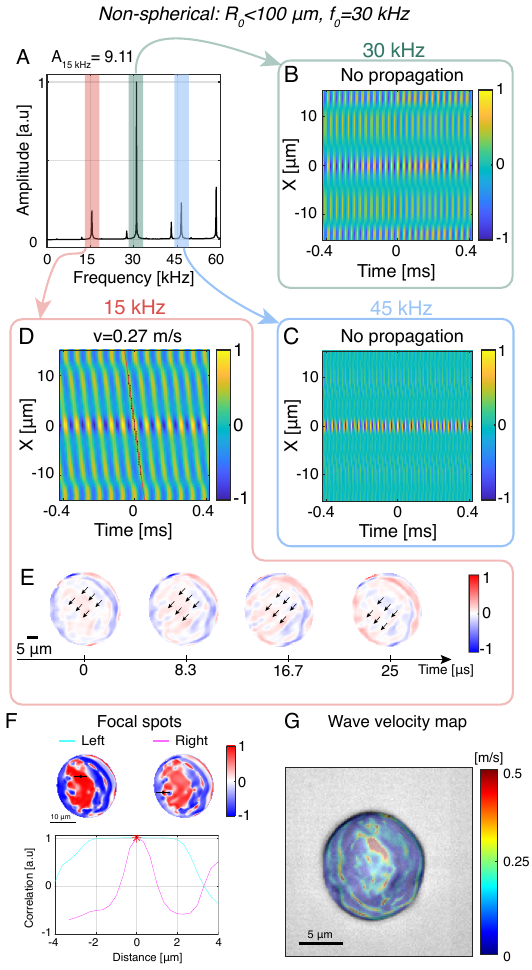}
	\caption{Shear wave velocity measurement inside the cell. (A) Spectrum of the displacement field. Average spatio-temporal correlation matrices at (B) \qty{30}{\kilo\hertz}, (C) \qty{45}{\kilo\hertz} and (D) \qty{15}{\kilo\hertz}. (E) Snapshots of the displacement field filtered at \qty{15}{\kilo\hertz}. (F) Two correlation focal spots calculated at different points marked by *, and their corresponding lateral profiles. (G) Final wave velocity map estimated inside the cell.}
	\label{fig3}
\end{figure}

Once elastic waves were generated and detected inside the cell, their velocity could be measured to mechanically characterize the cell. First, on the average correlation matrix displayed in fig.~\ref{fig3}D, the average shear wave velocity could be estimated. Indeed, the slope of the propagating patterns corresponds to the average wave velocity inside the cell. In this example, the average wave velocity was of \qty{0.27}{\metre\per\second}. Over 18 similar experiments performed on megakaryocytes involving nonspherically oscillating bubbles (exhibiting modes of degrees 3 to 6), the measured average wave velocity values were \qty[parse-numbers=false]{0.12 \pm 0.08}{\metre\per\second}. They are consistent with previously reported micro-elastography experiments on other cell lines \cite{grasland-mongrain_ultrafast_2018,fle_imaging_2023}. However, there is no ground truth to compare with for such a dynamic elasticity measurement performed on megakaryocytes.

Furthermore, using noise correlation algorithms, we could also reconstruct shear wave velocity maps. For this configuration involving nonspherically oscillating bubbles, exemplary correlation focal spots at two different locations are shown in fig.~\ref{fig3}F, and the corresponding wave velocity map is presented in fig.\ref{fig3}G. It is in good agreement with the average wave velocity measured in fig.\ref{fig3}D. It shows a wide range of wave velocities inside the cell, from 0.1 to \qty{1.3}{\metre\per\second}. 

At this stage, we have been able to take advantage of nonspherical oscillations of gas bubbles placed in the vicinity of a cell to induce shear waves inside the cell, that could be detected and tracked. Indeed, the nonspherical oscillations were crucial to generate the subharmonic frequency at \qty{15.4}{\kilo\hertz} allowing shear wave propagation. However, we could wonder whether nonspherical oscillations are actually needed, and if purely radial oscillations could be employed for cell micro-elastography. 

\subsection{Are nonspherical oscillations required?}
\begin{figure}[!t]
	\centering
	\includegraphics[scale=0.9]{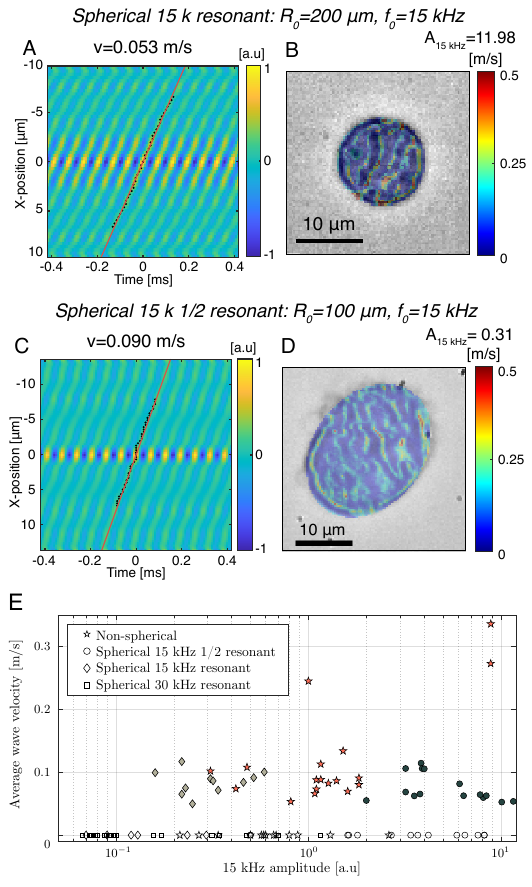}
	\caption{Other configurations that have been tested, varying the oscillation type, the bubble radius and the excitation frequency. For each configuration, the average spatio-temporal correlation matrices (A and C) and the wave velocity maps (B and D) are displayed. (E) Graphics summarizing all the experiments that have been made in the context of this paper: the 4 bubble oscillation conditions are represented with different symbols, and plotted as a function of the 15 kHz peak amplitude measured in the cell displacements spectrum and the measured wave speed in the cell. When a wave speed could not be measured, the marker is left white and put on the 0~m/s axis. }
	\label{fig4}
\end{figure}

In further experiments, we explored the possibility of generating mechanical waves propagating in cells via spherically oscillating bubbles. In fig.~\ref{fig4} are shown results for the two other bubble oscillations types that were employed to generate vibrations at \qty{15}{\kilo\hertz}. The most straightforward method to produce \qty{15}{\kilo\hertz} vibrations using an oscillating bubble is to employ a bubble of radius \qty{200}{\micro\metre} and excite it at its resonance frequency (\qty{15}{\kilo\hertz}). This configuration leads to bubble dynamics characterized by an oscillation around the equilibrium radius at \qty{30}{\kilo\hertz}, modulated by a \qty{15}{\kilo\hertz} component (see Fig.~S1A). The \qty{30}{\kilo\hertz} component is likely due to the transducer transfer function. This dual-frequency behavior is reflected in the displacement field spectrum measured inside the cell, which exhibits two peaks at \qty{15}{\kilo\hertz} and \qty{30}{\kilo\hertz} with comparable amplitudes (see Fig.S1B). The 15 kHz-peak amplitude is comparable to the nonspherical oscillation case (see $\mathrm{A_{15 kHz}}$ in fig.~\ref{fig4}). The average spatio-temporal correlation matrix of the displacement field filtered around \qty{15}{\kilo\hertz} is displayed in Fig.~\ref{fig4}A, showing clear oblique patterns corresponding to propagating waves. The slope was found to be \qty{0.053}{\metre\per\second}, which corresponds to the average shear wave velocity inside the cell. Furthermore, a shear wave velocity map could be extracted and is shown in Fig.~\ref{fig4}B. The wave velocity values that were measured are compatible with the average wave velocity and show high velocity contrast in the cell. These results confirm that resonant bubbles excited at \qty{15}{\kilo\hertz} are suitable for micro-elastography. However, from the experimental point of view, having a cell and a bubble of the right size so close from each other without killing the cell or having the bubble exploding or imploding was difficult to achieve. Indeed, the bubble size had to be well controlled in order to get a bubble that was resonant at \qty{15}{\kilo\hertz} and ; in addition, the large size mismatch between the bubble and the cell (approximately 20-fold) made it challenging to position the bubble close to the cell without compromising cell viability.

To overcome this limitation, we investigated an alternative approach to generate \qty{15}{\kilo\hertz} vibrations. In this configuration, bubbles of radius around \qty{100}{\micro\metre} (resonant at \qty{30}{\kilo\hertz}) were excited at half their resonance frequency. In this case, the bubble oscillation consisted of \qty{30}{\kilo\hertz} spherical oscillations modulated in amplitude by a \qty{15}{\kilo\hertz} component (see Fig.~S2A). Similarly to the previous configuration, the displacement field spectrum inside the cell showed two components at 15 and \qty{30}{\kilo\hertz} (see Fig.S2B), although the \qty{15}{\kilo\hertz} peak was approximately ten times weaker than the \qty{30}{\kilo\hertz} peak. By isolating the \qty{15}{\kilo\hertz} component, the average spatio-temporal correlation matrix was computed and is shown in Fig.~\ref{fig4}C. Clear evidence of wave propagation was observed, and the slope of the oblique patterns yielded an average wave velocity of \qty{0.090}{\metre\per\second}. The corresponding shear wave velocity map is presented in Fig.~\ref{fig4}D.

Lastly, as a control experiment, we tested bubbles resonant at \qty{30}{\kilo\hertz} excited at their resonance frequency. In this case, the displacement spectrum exhibited a single peak at \qty{30}{\kilo\hertz}, and no wave propagation was observed in the spatio-temporal correlation matrix (see Fig.S4). This principle is illustrated in Fig.~\ref{fig4}E, which summarizes all experiments performed on megakaryocytes. The four bubble oscillation conditions are represented: non-spherical oscillations of bubbles of any size (Fig.~\ref{fig2}), spherical oscillations of large bubbles at \qty{15}{\kilo\hertz} (Fig.~\ref{fig4}A-B), spherical oscillations of smaller bubbles at half their resonance frequency (Fig.~\ref{fig4}C-D) and spherical oscillations at \qty{30}{\kilo\hertz} (Fig.S4). Each experiment is plotted in a 2D space showing the measured average wave velocity and the amplitude of the \qty{15}{\kilo\hertz} peak in the displacement spectrum. If no wave velocity could be measured, the velocity is set to \qty{0}{\metre\per\second} and the marker is left white. These findings reinforce the conclusion that the presence of a \qty{15}{\kilo\hertz} component in the bubble oscillation is essential for micro-elastography, regardless of whether the oscillation is spherical or non-spherical. However, if the bubble oscillation lacks a \qty{15}{\kilo\hertz} component and all frequencies are above the cutoff frequency, no wave propagation occurs and micro-elastography becomes impossible. In summary, micro-elastography was successful in all configurations that generated \qty{15}{\kilo\hertz} vibrations, while no wave propagation was observed for pure \qty{30}{\kilo\hertz} oscillations. These results demonstrate that an oscillating microbubble placed near a cell is an effective vibration source for micro-elastography, provided its frequency content is compatible with the viscoelastic properties of the cell and does not exceed the cutoff frequency.

\section{Discussion}
The results exposed in this work show that the use of an oscillating bubble between 80 and \qty{400}{\micro\metre} in diameter makes possible contactless micro-elastography on cells of approximately \qty{15}{\micro\metre}. Contrary to previous studies \cite{grasland-mongrain_ultrafast_2018,fle_imaging_2023}, the absence of micro-pipette avoids the presence of artifacts on the wave velocity maps and does not affect the cell in any way, mechanically or in terms of viability. Other studies studying cell deformation induced by an oscillating bubble were destructive for the cell \cite{tandiono_resonant_2013} since bubble collapse was involved. Moreover, since there is no need to hold the cell with one micro-pipette and vibrate it with a second micro-pipette, we can characterize cells of any size. Indeed, so far, micro-elastography studies were limited to oocytes which are exceptionally large cells (approximately \qty{100}{\micro\metre}). In this study, we managed to perform measurements on cells up to 10 times smaller (10 to \qty{25}{\micro\metre} in diameter). Finally, the method allows to obtain a full wave velocity map in less than a millisecond, which could allow to follow very fast biological processes, contrary to other techniques relying on scanning such as Atomic Force Microscopy or indentation.

Most advantages of this technique are due to the use of an original shear wave source which is an oscillating microbubble placed in the vicinity of the cell. In the past, either bubble generation \cite{montalescot_electrolysis-induced_2016} or bubble collapse \cite{izak_ghasemian_shear_2023} have been used for generating shear waves. Here, we use stable oscillating bubbles as elastic wave sources. We showed that the method is robust and that several bubble oscillation conditions can be used, but did not discuss the wave generation mechanism so far. First, we can wonder why a bubble is needed and why the ultrasonic wave generated by the transducer itself can not be used as a shear wave source directly. This configuration was tested as a control experiment and no waves could be observed inside the cell. Indeed, the bubble can be seen as a local particle velocity amplifier for the ultrasonic field. The bubble, in reaction to the ultrasonic field generated by the transducer, oscillates and re-radiates energy in the medium in the form of compression waves. The bubble can then be seen as a point source for ultrasound. However, it does not only re-radiates spherical waves in the surrounding medium, but it acts as a velocity field amplifier. Indeed, for the example shown in Fig.~\ref{fig2} where a bubble with an equilibrium radius of \qty{117.4}{\micro\metre} exhibits radial oscillations of amplitude \qty{2}{\micro\metre}, assuming linear radial oscillation \cite{fauconnier_nonspherical_2020} allows estimating the local driving pressure around the bubble, which was \qty{1.3}{\kilo\pascal}. On one hand, the particle velocity amplitude associated to this ultrasound wave was \qty{1.1}{\milli\metre\per\second}, using the acoustical impedance of water. On the other hand, we could estimate the particle velocity amplitude of the wave radiated by the bubble at a \qty{7}{\micro\metre} distance \cite{doinikov_acoustic_2019}, which was \qty{30.4}{\centi\metre\per\second}. Hence, the oscillating bubble behaves as an efficient particle velocity amplifier for objects placed in its vicinity, with a gain of 300 in this example, which generates propagating elastic waves inside the cell. If the bubble oscillates non-spherically, the geometry of the bubble deformation generates shear stress at the surface of the cell. Alternatively, if it oscillates spherically, the presence of the cell breaks the spherical symmetry and ultimately generates some shear strain. Hence, both non-spherically and spherically oscillating bubbles are suitable for generating shear waves inside cells.

However, as a last discussion point, we can wonder which bubble oscillation type is the most efficient or appropriate for this application. Several configurations have been tested for the bubble oscillation, and 3 out of 4 were shown to be usable for micro-elastography. Indeed, as far as there was a \qty{15}{\kilo\hertz}-component in the displacement field, whether the bubble oscillated spherically or not, propagating waves could be detected inside the cells. However, in terms of experimental feasibility, not all configurations are similar. Firstly, the configuration performing best in terms of 15 kHz vibrations amplitude inside the cell is the configuration called ``Spherical 15 k resonant", but working with bubbles approximately 20 times bigger than the cells (\qty{400}{\micro\metre} vs \qty{20}{\micro\metre} in diameter) is particularly challenging. The risk of killing the cell while manipulating the gas bubble in its surroundings is very high, and the positioning of the bubble is not easy to achieve. Secondly, since it uses smaller bubbles, the configuration ``Spherical 15k 1/2 resonant" is easier to achieve but the \qty{15}{\kilo\hertz}-component remains quite low. In addition, the bubble radius needs to be very well controlled in order for the bubble to be excited exactly at the subharmonic of its resonance frequency. Finally, the nonspherical configuration presents many advantages: bubbles as small as \qty{65}{\micro\metre} and up to more than \qty{200}{\micro\metre} of radius can be employed, reducing the need for a good control over the bubble radius. In this configuration, whatever the bubble size, exceeding the pressure threshold for parametric instability allows triggering nonspherical oscillations. Hence, the subharmonic \qty{15}{\kilo\hertz} vibration is efficiently transmitted towards the inside of the cell. Therefore, in the end, nonspherically oscillating bubbles are more convenient to be used as wave sources for micro-elastography since a large bubble radius range can be used. This method could be further extended to a more systematic experiment where cells could be characterized one-by-one as they pass in front of a bubble, introducing the ``elastographic cytometry".

\section{Conclusion}
We showed that employing an oscillating bubble as a shear wave source unlocks contactless micro-elastography. Megakaryocytes as small as \qty{10}{\micro\metre} in diameter could be characterized without the need for micro-pipettes and hence no contact artifacts. To our knowledge, this is the first report of micro-elastography on cells much smaller than often studied oocytes. Both spherical and nonspherical bubble oscillations were tested. They were found to be suitable for micro-elastography purposes provided the spectrum of the oscillation contains low enough frequencies.

\matmethods{
\subsection{Experimental setup}
The experimental setup used was similar to the one previously used for studying the dynamics and microstreaming induced by a nonspherically substrate-attached microbubble \cite{fauconnier_nonspherical_2020,fauconnier_nonspherical_2022}. It was composed of a tank (\SI{250\times100\times40}{\cubic\milli\meter}) filled with cell culture medium (Dulbecco’s Modified Eagle medium DMEM, high glucose, without L-glutamin). On the tank bottom were placed a bubble and a cell at a distance ranging between 5 and \SI{30}{\micro\meter}. The cells were megakaryocytes (cell line MEG-01, CRL-2021, American Type Culture Collection, Manassas, Virginia) which are blood cells producing the platelets. Their diameter varied between 10 and \SI{100}{\micro\meter}, but the cell size distribution analysis performed over 500 randomly hand-picked cells exhibited a peak near \qty{16}{\micro\metre} in diameter. A dihydrogen microbubble was created at the tip of the cathode of an electrolysis actuator. The cathode connector was tied up to a three-axis hydraulic micro-manipulator which allowed the positioning and tethering of the bubble at the tank's bottom. Once the bubble was stable on the bottom of the tank, the electrode was moved away so that it did not interfere with the oscillations of the attached bubble. The bubble radius was ranged between 50 and \SI{200}{\micro\meter}. The tank was placed under an inverted microscope (Nikon Ti2-U) with a $\times$60 magnification lens, connected to a high-frame-rate camera (Vision Research, Phantom V12.1). For each experiment, two videos were acquired: a low-frame rate video with a wide field-of-view allowing to see at least part of the bubble and the cell, and a higher frame rate video showing only the cell(see table~\ref{table} for details). The low frame rate video was used to analyze bubble dynamics, but since the frame rate was too low to resolve the dynamics, we used it in a stroboscopic mode and refolded the data on 2 acoustic periods. A Langevin-type ultrasound transducer (Sinaptec, Lezennes, France, 30.8kHz resonance frequency) was immersed in the tank and driven by a continuous excitation signal at its resonance frequency or the subharmonic frequency (\SI{15.4}{\kilo\hertz}). The applied acoustic pressures were quantified as follows. For a given applied pressure, radial oscillations of the bubble were recorded and fitted using a linearized Rayleigh-Plesset modeling. As the fluid viscosity and density, and the bubble equilibrium radius were known, the acoustic pressure could be deduced from this modeling. This technique was reproduced for increasing driving voltage as far as the bubble interface remained spherical. Acoustic pressures associated to bubble nonspherical oscillations were extrapolated linearly along the whole experimental data set. We deduced that applied acoustic pressures ranged between 0 and 40~kPa.

\begin{table}[ht]
	\begin{center}
		\begin{tabular}{c|c|c|c}
			\textbf{Image type} & \textbf{Image size} & F\textbf{Frame rate} & \textbf{Exposure} \\ \hline
			Bubble+cell & 800$\times$800 & 6.27 kHz & \SI{7.9}{\micro\second} \\
			Cell only & 128$\times$128 & 120.1 kHz & \SI{8.3}{\micro\second} \\
		\end{tabular}
			\end{center}
		\caption{Acquisition parameters for the two different types of videos acquired for each experiment. }
		\label{table}
	\end{table}

\subsection{Cell culture}
Cells were grown in 75 cm² flasks (Corning) filled with 15mL Roswell Park Memorial Institute (RPMI) medium supplemented with 10\% fetal calf serum (FCS) and 1\% L-glutamin, then incubated at 37°C and 5\% CO2. The medium renewal was done twice a week.
	
\subsection{Bubble oscillations analysis}
The bubble oscillations were driven at a frequency of \qty[parse-numbers=false]{f=30.8}{\kilo\hertz}, which corresponds to the resonant frequency of the Langevin transducer, except for the configuration ``Spherical 15k 1/2 resonant" where the bubbles were driven at the subharmonic frequency (\qty{15.4}{\kilo\hertz}). Whatever the applied acoustic pressure, the bubble oscillates spherically, linearly or not. However, only above a given pressure threshold, the bubble behavior might change from purely radial to nonspherical oscillations \cite{leighton_acoustic_1994}, triggered by parametric instabilities.

The bubble modal content is classically described by the set of well-known spherical harmonics of degree n and order m (where $-n \leq m \leq n$), although providing an exact quantification can be difficult in the presence of asymmetric modes – promoted by bubble attachment to a substrate \cite{fauconnier_nonspherical_2020,cattaneo_cyclic_2025} – visualized from a single (top) view only. Conveniently, because the pressure threshold and the modal shape are function of the bubble equilibrium radius and the driving frequency \cite{francescutto_pulsation_1978}, the modal degree can be retrieved based on the bubble size information only \cite{fauconnier_nonspherical_2020}. This assumes that the surface oscillation was triggered on the first parametric resonance, as given by the Lamb spectrum \cite{lamb_hydrodyanmics_1932}.

In this work, selecting resonant bubbles, which preferably oscillate on sectoral modes ($n=m$) at low acoustic amplitude \cite{fauconnier_nonspherical_2020}, facilitates the analysis of their modal content, as such sectoral deformations can be exactly quantified from a spatial Fourier transform of the bubble top-view contour \cite{fauconnier_nonspherical_2020}. The obtained modal coefficients are then weighted based on the known analytical expression of the spherical harmonics, and its top-viewed projected contour, as detailed in a previous work \cite{fauconnier_nonspherical_2020}. Throughout this work, the focus was on sectoral modes of degree n = 3 to 6, referred to as “modes of degree 3 to 6”.

\subsection{Displacements estimation}
The displacement field inside the cell was calculated using a phase-tracking algorithm already used in previous works \cite{laloy-borgna_micro-elastography_2021,delattre_passive_2021,gregoire_flexural_2024}. It relies on the phase difference calculation of the Hilbert transformation of optical images. It can be employed to measure the displacement field in both directions of an image, with a sub-pixel sensitivity.
	
\subsection{Noise correlation elastography} 
Noise correlation elastography algorithms were used to retrieve shear wave velocity from the displacement field \cite{catheline_tomography_2013}. They rely on temporal cross-correlation of displacement signals measured at different pixels. They can be used to measure both the average wave velocity over the whole cell, or to reconstruct shear wave velocity maps. First, to measure the average wave velocity in the cell, the average spatio-temporal correlation matrix was calculated, like already done in previous works \cite{laloy-borgna_observation_2023,gregoire_flexural_2024}. x being a spatial coordinate, t the time, $\phi$ the displacement field and $\oplus$ the temporal correlation, the average spatio-temporal correlation matrix in one direction can be written as:
\begin{equation}
	\mathcal{T}o\mathcal{F}(\Delta x, \Delta t) = \langle \phi(x_0,t) \oplus_{\Delta t} \phi(x_0+\Delta x,t)\rangle _{x_0}
\end{equation}
	
Then, a linear fit on the maxima of such a matrix allowed the measurement of the average wave velocity inside the cell (see fig.\ref{fig3}D).
	
To reconstruct wave velocity maps, temporal cross-correlation of the displacement field had to be calculated. At each pixel, temporal correlation of the displacement signal with displacement signals of neighboring pixels was computed, resulting in a so-called called focal spot \cite{grasland-mongrain_ultrafast_2018,zemzemi_super-resolution_2020} (see fig.~\ref{fig3}F). The size and curvature of this focal spot is linked to the local shear wavelength at this specific pixel. By repeating the process for all pixels, a shear wavelength map can be obtained. Independently, the local central frequency at each pixel was estimated, finally resulting in wave velocity maps.
}

\showmatmethods{} 

\acknow{The authors warmly acknowledge Magali Perier and Jacqueline Ngo for their valuable help with cell culture. This work was supported by a grant from the French National Research Agency as part of the ``Investissements d’Avenir ExcellencES" program from France 2030 (SHAPE-Med@Lyon ; ANR-22-EXES-0012). M.F would like to thank the graduate school MEGA and Labex CeLyA for their support.}

\showacknow{} 

\bibsplit[15]

\bibliography{Biblio}

\end{document}